\documentclass[prd,superscriptaddress,showpacs,showkeys]{revtex4}
\usepackage{bm}
\usepackage{amsmath}
\usepackage{color}
\usepackage{ulem}

\newcommand{\tr}{\mbox{tr}}
\newcommand{\sign}{\mbox{sign}}

\begin{document}

\title{Chiral asymmetry and axial anomaly in magnetized relativistic matter}

\date{October 29, 2010}

\author{E. V. Gorbar}
\email{gorbar@bitp.kiev.ua}
\affiliation{Bogolyubov Institute for Theoretical Physics, 03680, Kiev, Ukraine}

\author{V. A. Miransky}
\email{vmiransk@uwo.ca}
\affiliation{Department of Applied Mathematics, University of Western Ontario, London, Ontario N6A 5B7, Canada}

\author{I. A. Shovkovy}
\email{igor.shovkovy@asu.edu}
\affiliation{Department of Applied Sciences and Mathematics, Arizona State University, Mesa, Arizona 85212, USA}

\begin{abstract}
The induced axial current and the chiral anomaly are studied in the normal phase of magnetized
relativistic matter. A special attention is paid to the role of the chiral shift parameter $\Delta$, leading
to a relative shift of the longitudinal momenta in the dispersion relations of opposite chirality fermions.
In the Nambu-Jona-Lasinio model, it is shown directly from the {\it form} of the gap equation that 
$\Delta$ necessarily exists in the normal phase in a magnetic field. By making use of the gauge invariant
point-splitting regularization, we then show that the presence of  $\Delta$ essentially modifies
the form of the axial current, but does not affect the conventional axial anomaly relation. By recalculating
the axial current with the proper-time regularization, we conclude that the result is robust with respect to
a specific regularization scheme used.
\end{abstract}


%  12.39.Ki Relativistic quark model
%  12.38.Mh Quark-gluon plasma
%  21.65.Qr Quark matter
%  26.60.Dd Neutron star core
%  25.75.Nq Quark deconfinement, quark-gluon plasma production, and phase transitions


\maketitle

\section{Introduction}
\label{1}

Several types of dense relativistic matter exist in compact stars. For example, a relativistic electron
plasma forms and plays an essential role in white dwarfs. Also, electrons form a relativistic fluid inside
nuclear matter in the interior of neutron stars. If quark stars exist in nature, the corresponding dense
quark matter in the core will be a strongly coupled version of relativistic matter. Often, such matter is
subject to strong magnetic fields. In white dwarfs, e.g., the magnetic fields reach up to $10^{9}~\mbox{G}$,
while, in neutron stars, they may be up to $10^{15}~\mbox{G}$ \cite{astroreview,astroreview1}.
Relativistic matter in a strong magnetic field is also created in heavy ion collisions \cite{Skokov:2009qp}
that can lead to the chiral magnetic effect \cite{Kharzeev:2007tn}.

Many physical properties of the stellar matter under extreme conditions realized inside compact
stars are understood theoretically and could be tested to some extent through observational
data. However, as was pointed out in Refs.~\cite{FI1,Metlitski:2005pr,Gorbar:2009bm,Rebhan,
Basar:2010zd,Fukushima,Kim,Frolov}, the dense relativistic matter in a strong magnetic field may
hold some new theoretical surprises. In particular, a topological contribution in the axial current
at the lowest Landau level (LLL) was revealed in Ref.~\cite{Metlitski:2005pr}. More recently,
it was shown in Ref.~\cite{Gorbar:2009bm}, that in the normal phase of dense relativistic matter
in a magnetic field, there exists a contribution to the axial current associated with a relative shift
of the longitudinal momenta in the dispersion relations of opposite chirality fermions,
$k^3 \to k^3 \pm \Delta$, where the momentum $k^3$ is directed along magnetic field and
$\Delta$ is the chiral shift parameter intimately connected with the induced axial current
$j^{3}_{5}$. Unlike the topological contribution in $j^{3}_{5}$ at the lowest Landau level
(LLL) \cite{Metlitski:2005pr}, the dynamical one appears only in interacting matter and affects
the fermions in {\it all} Landau levels, including those around the Fermi surface. The induced axial
current and the shift of the Fermi surfaces of the left-handed and right-handed fermions are expected
to play an important role in transport and emission properties of matter in various types of
compact stars as well as in heavy ion collisions.

The main goal of this Letter is to study some general and subtle features of the dynamics with the
chiral shift parameter $\Delta$. One of such issues is whether the form of the induced axial $j^{3}_{5}$
current coincides with the result in the theory of noninteracting fermions in a magnetic field
\cite{Metlitski:2005pr} or whether it is affected by interactions (for related discussions, see
Refs.~\cite{Metlitski:2005pr,Gorbar:2009bm,Rebhan,Fukushima,Rubakov}). This question is
intimately related to that of the connection of the structure of the induced $j^{3}_{5}$ with 
the axial anomaly \cite{ABJ}. By using the Nambu-Jona-Lasinio (NJL) model in a magnetic field,
it will be shown that while the dynamics responsible for the
generation of the chiral shift $\Delta$ essentially modifies the form of this current, it does {\it not}
affect the form of the axial anomaly: the latter is connected only with the topological part in the LLL 
\cite{Ambjorn}. Moreover, while the topological contribution in the axial
current is generated in the infrared kinematic region (at the LLL), the contribution of $\Delta$
in this current is mostly generated in ultraviolet, which implies that higher Landau levels
are primarily important in that case.

\section{Model: General properties}
\label{2}

As in Ref.~\cite{Gorbar:2009bm}, in order to illustrate this phenomenon in the clearest way,
we will utilize the simplest NJL model with one fermion flavor, whose
Lagrangian density is
\begin{eqnarray}
{\cal L} &=& \bar\psi \left(iD_\nu+\mu_0\delta_{\nu}^{0}\right)\gamma^\nu \psi
-m_{0}\bar\psi \psi  +\frac{G_{\rm int}}{2}\left[\left(\bar\psi \psi\right)^2
+\left(\bar\psi i\gamma^5\psi\right)^2\right],
\label{NJLmodel}
\end{eqnarray}
where $m_{0}$ is the bare fermion mass and $\mu_0$ is the chemical potential. By
definition, $\gamma^5\equiv i\gamma^0\gamma^1\gamma^2\gamma^3$. The covariant
derivative $D_{\nu}=\partial_\nu + i e A_{\nu}$ includes the external gauge field
$A_{\nu}$, which is assumed to be in the Landau gauge, $A^{\nu}= x B \delta_{2}^{\nu}$
\cite{footnote1}. Here $B$ is the strength of the external magnetic field pointing in the
$z$-direction. The $(3+1)$-dimensional Lorentz symmetry in the model is explicitly broken down
to the $SO(2)$ symmetry of rotations around the $z$-axis in the presence of this magnetic field.
Also, except parity ${\cal P}$, all the discrete symmetries ${\cal C}$,  ${\cal T}$,
${\cal CP}$, ${\cal CT}$, $PT$, and ${\cal CPT}$ are broken (here ${\cal C}$
and ${\cal T}$ are charge conjugation and time reversal, respectively).

In the chiral limit, $m_{0}=0$, this model possesses the chiral $U(1)_L\times U(1)_R$
symmetry. In the vacuum state ($\mu_0=0$), however, this chiral symmetry is known to be
spontaneously broken because of the magnetic catalysis phenomenon \cite{MC}. In essence,
such spontaneous breaking results from the enhanced pairing dynamics of fermions and
antifermions in the infrared. The enhancement results from the non-vanishing density of
states in the lowest Landau level that is subject to an effective dimensional reduction
$D\to D-2$. At a sufficiently large value of the chemical potential, the chiral symmetry is
expected to be restored. As we shall see below, this is indeed the case, but the corresponding
normal ground state is characterized by a nonzero chiral shift parameter $\Delta$.

We will analyze model (\ref{NJLmodel}) in the mean field approximation, 
which is reliable in the weakly coupled regime when the dimensionless coupling constant
$g \equiv G_{\rm int}\Lambda^2/(4\pi^2) \ll 1$, where $\Lambda$ is an ultraviolet cutoff.
Note that here the coupling $g$ is defined in such a way that $g_{cr} =1$, where $g_{cr}$ is
the critical value for generating a fermion dynamical mass in the NJL model without 
magnetic field.

In this approximation, the full fermion propagator does not allow any wave function
renormalization different from $1$. Thus, the general ansatz for the (inverse) full propagator
is given by \cite{Gorbar:2009bm}
\begin{eqnarray}
iG^{-1}(u,u^\prime) &=&\Big[(i\partial_t+\mu)\gamma^0 -
(\bm{\pi}\cdot\bm{\gamma})
+ i\tilde{\mu}\gamma^1\gamma^2
+\Delta\gamma^3\gamma^5
-m\Big]\delta^{4}(u- u^\prime),
\label{ginverse}
\end{eqnarray}
where $u=(t,\bm{r})$, $\pi^k = i (\partial^k + i e A^{k})$ is the canonical
momentum, $m$ is the constituent (medium-modified) fermion mass,
$\mu$ is an effective chemical potential in
the quasiparticle dispersion relation, $\tilde{\mu}$ is an anomalous  magnetic moment, and
$\Delta$ is the chiral shift parameter. Note that $\mu$ in the full propagator may differ from
the thermodynamical chemical potential $\mu_0$ in Lagrangian density (see below).
As is shown in Appendix \ref{AppPropagator}, in the mean field approximation one has
$\tilde{\mu} = 0$ in a self-consistent solution to the gap equation in this model.

Let us now demonstrate that one can get an important insight into the properties of the
solutions in this model already from the {\it form} of the gap equation for the parameters
$\mu$, $\Delta$, and $m$. As described in more detail in Appendix \ref{AppPropagator},
utilizing the approach based on the effective action for composite operators \cite{CJT},
one can show that the gap equation in the mean field approximation reduces to the
following set of equations:
\begin{eqnarray}
\mu -  \mu_0 &=& -\frac{1}{2}G_{\rm int} \langle j^0\rangle ,
\label{gap-mu}  \\
\Delta &=& -\frac{1}{2}G_{\rm int} \langle j_5^3\rangle ,
\label{gap-Delta} \\
m - m_0 &=&  -G_{\rm int} \langle \bar\psi \psi\rangle ,
\label{gap-m}
\end{eqnarray}
where the chiral condensate $\langle \bar\psi \psi\rangle$, the vacuum expectation values
of the fermion density $j^0$ and the axial current density $j_5^3$ are
\begin{eqnarray}
\langle j^0\rangle &=& -\tr\left[\gamma^0G(u,u)\right],
\label{charge}  \\
\langle j_5^3\rangle &=& -\tr\left[\gamma^3\gamma^5G(u,u)\right],
\label{current} \\
\langle \bar\psi \psi\rangle &=& -\tr\left[G(u,u)\right].
\label{condensate}
\end{eqnarray}
Let us now consider the case of the normal phase in the chiral limit, when $m = m_0 = 0$
and $\langle \bar\psi \psi\rangle = 0$. It is realized when the chemical potential
$\mu_0 > m_{dyn}/\sqrt{2}$ \cite{Gorbar:2009bm}, where $m_{dyn}$ is a dynamical
fermion mass in a magnetic filed at zero chemical potential and zero temperature. Let us
analyze Eqs.~(\ref{gap-mu}) and (\ref{gap-Delta}) in perturbation theory in the coupling
constant $g$. In the zero approximation, we have a theory of free fermions in a magnetic
field. To this order, $\mu=\mu_0$ and $\Delta=0$. However, even in this case the fermion
density $\langle j^0\rangle$ and the axial current density $\langle j_5^3\rangle$ are
nonzero. The former can be presented as a sum over the Landau levels:
\begin{eqnarray}
\langle j^0\rangle_0 &=& \frac{ \mu_0|eB|}{2\pi^2} +\frac{\sign(\mu_0)|eB|}{\pi^2}
\sum_{n=1}^{\infty}
\sqrt{\mu_{0}^2-2n|eB|}
 \theta\left(|\mu_0|-\sqrt{2n|eB|}\right),
\label{j}
\end{eqnarray}
and the latter is \cite{Metlitski:2005pr}
\begin{equation}
\langle j^3_5\rangle_0 = \frac{eB}{2\pi^2}\mu_0\, .
\label{MZ}
\end{equation}
Then, to the next order in the coupling constant, one finds from Eq.~(\ref{gap-Delta}) 
that $\Delta \propto G_{\rm int} \langle j^3_5\rangle_0 \neq 0$. Thus, in the normal phase of
this theory, there {\it necessarily} exists a nonzero shift parameter $\Delta$. In fact, this is 
one of the main results of Ref.~\cite{Gorbar:2009bm}. Let us also emphasize that $\Delta$ 
is generated by perturbative dynamics, which is directly connected with the fact that the vanishing 
$\Delta$ is not protected by any symmetry [recall that ${\cal C} =+1$, ${\cal P} =+1$, and 
${\cal T} = -1$ for the axial current density $j_5^3$, and beside parity ${\cal P}$, all the discrete 
symmetries are broken in model (\ref{NJLmodel})]. 

Similarly, one finds from Eq.~(\ref{gap-mu}) that 
$\mu - \mu_0 \propto G_{\rm int}\langle j^0\rangle_0 \neq 0$, i.e., $\mu$ and $\mu_0$ are different. 
One can trace the origin of this difference to the Hartree terms in the gap equation [see the last two 
terms in Eq.~(\ref{gap})]. This seems to be robust in the NJL model with a local four-fermion interaction 
and a chemical potential, associated with a global charge, such as a baryon (or lepton) charge for example. 
Note that when the conserved charge is related to a gauge symmetry, as in the case of the electric charge, 
the situation may be different. In that case, a neutrality condition imposed by the Gauss law 
takes place \cite{Kapusta}. The latter is necessary for providing the thermodynamic equilibrium 
in a system. This is likely to result in $\mu^{(e)}=\mu_{0}^{(e)}$ when $\mu^{(e)}$ is the chemical 
potential for electric charge. Note that usually there are chemical potentials of both types in dense
relativistic matter. While being of importance for potential applications in principle, we expect 
that this fact will not change our main conclusion regarding the chiral shift parameter.

By noting from Eq.~(\ref{gap-Delta}) that the chiral shift $\Delta$ is induced by the axial
current, it is naturally to ask whether $\Delta$ itself affects this current. The answer to 
this question is affirmative \cite{Gorbar:2009bm}. Another natural question is whether 
the divergence of this modified current satisfies the conventional anomaly equation 
\cite{ABJ}. As will be shown in the next section, the answer to this question is also 
affirmative.

\section{Induced axial current and axial anomaly}
\label{3}

In this section, using the gauge invariant point-splitting regularization, we study the influence
of the shift parameter $\Delta$ on the form of the axial current and the axial anomaly (for
reviews of this regularization, see Refs.~\cite{Peskin,Ioffe}). The analysis is model independent
and is based only on the form of the fermion propagator with $\Delta$ in an external
electromagnetic field. Our main conclusion is that while including $\Delta$ essentially
changes the form of the axial current, it does not modify the axial anomaly. Moreover, while
the contribution of the chemical potential in the axial current is generated in the infrared kinematic
region (at the LLL) \cite{Metlitski:2005pr}, the contribution of $\Delta$ in the current is
mostly generated in ultraviolet (at all Landau levels).

\subsection{Axial current}
\label{3a}

We consider the case of a constant electromagnetic field. Since it is known that the axial anomaly
is insensitive to chemical potential \cite{Son,Gavai}, the latter will be omitted. Then, the general
form of the fermion propagator is \cite{Schwinger}
\begin{equation}
G(u,u^{\prime}) = e^{i\Phi(u,u^{\prime})}\bar{G}(u - u^{\prime})
\end{equation}
with the Schwinger phase
\begin{equation}
\Phi(u,u^{\prime})=e\int_{u^{\prime}}^u dx^{\nu}A_{\nu},
\label{phase}
\end{equation}
where the integration is performed along the straight line. The translation invariant part
$\bar{G}(u - u^{\prime})$ depends only on the field strength $F_{\mu\nu}$.

Therefore, in the normal phase with $m=0$, the inverse propagator (\ref{ginverse})
(with $\tilde{\mu}=0$) can be rewritten as
\begin{eqnarray}
iG^{-1}&=&iD_{\nu}\gamma^{\nu}+\Delta\gamma^3\gamma^5
=\left(iD_{\nu}\gamma^{\nu}-\Delta s_\perp\gamma^3\right) {\cal P}^{-}_5+
\left(iD_{\nu}\gamma^{\nu}+\Delta s_\perp\gamma^3\right) {\cal P}^{+}_5\,,
\end{eqnarray}
where $s_\perp \equiv \sign(eB)$, $D_{\nu}=\partial_{\nu}+ ieA_{\nu}$, and
${\cal P}^{\mp}_5=(1 \mp  s_\perp \gamma^5)/2$. This equation implies that the
effective electromagnetic vector potential equals $\tilde{A}_{\nu}^{-}=A_{\nu}
+ (s_\perp \Delta /e)\delta_{\nu}^{3} $ and $\tilde{A}_{\nu}^{+}=A_{\nu} -
(s_\perp \Delta /e)\delta_{\nu}^{3}$ for the $-$ and $+$ chiral fermions,
respectively. Since the field strength $F_{\mu\nu}$ for $\tilde{A}_{\nu}^{\mp}$
is the same as for $A_{\nu}$, $\Delta$ affects only the Schwinger phase
(\ref{phase}):
\begin{eqnarray}
\Phi^{-}_{\Delta}(u,u^{\prime}) = \Phi(u,u^{\prime}) + s_{\perp}\Delta (u^3-u^{\prime\,3}),\\
\Phi^{+}_{\Delta}(u,u^{\prime}) = \Phi(u,u^{\prime}) - s_{\perp}\Delta
(u^3-u^{\prime\,3}).
\end{eqnarray}
Thus, we find
\begin{eqnarray}
G(u,u^{\prime})&=&
  \exp[is_{\perp}\Delta (u^3-u^{\prime\,3})]\,{\cal P}^{-}_5\,G_0(u,u^{\prime})
+\exp[-i s_{\perp}\Delta (u^3-u^{\prime\,3})]\,{\cal
P}^{+}_5\,G_0(u,u^{\prime})\,, \label{propagator-transformed}
\end{eqnarray}
where $G_0$ is the propagator with $\Delta=0$. Note that $\Delta$ appears now only in the phase factors.

According to Eq.~(\ref{current}), the axial current density is equal to
\begin{equation}
\langle j^{\mu}_5(u)\rangle=-\tr
\left[\gamma^{\mu}\gamma^5\,G(u,u+\epsilon)\right]_{\epsilon \to 0}. 
\label{density1}
\end{equation}
On the other hand, the fermion propagator in an electromagnetic field has the
following singular behavior for $u^{\prime}-u=\epsilon \to 0$
\cite{Peskin,Ioffe}:
\begin{equation}
G_0(u,u+\epsilon)\simeq
\frac{i}{2\pi^2}\left[\frac{\hat{\epsilon}}{\epsilon ^4} 
-\frac{1}{16\epsilon ^2}eF_{\mu\nu} \left(\hat{\epsilon}\sigma^{\mu\nu}
+\sigma^{\mu\nu}\hat{\epsilon } \right)\right], \label{asymptotics}
\end{equation}
where $\hat{\epsilon }=\gamma_{\mu}\epsilon^{\mu}$. Then using
Eqs.~(\ref{propagator-transformed}) -- (\ref{asymptotics}), we find
\begin{eqnarray}
\langle j^{\mu}_5 \rangle_{\rm singular} &=& \left.
\frac{i\epsilon^{\mu}s_\perp}{\pi^2\epsilon^4} \left(e^{- is_\perp
\Delta \epsilon^3}- e^{is_\perp  \Delta \epsilon^3}\right) +
\frac{ieF_{\lambda\sigma}\epsilon_{\beta}\epsilon^{\beta\mu\lambda\sigma}}
{8\pi^2\epsilon^2} \left(e^{- i s_\perp \Delta \epsilon^3}+e^{i
s_\perp \Delta \epsilon^3}\right)\right|_{\epsilon \to 0}\,.
\label{axialcurrent}
\end{eqnarray}
Taking into account that the limit $\epsilon \to 0$ should be taken in this
equation symmetrically \cite{Peskin,Ioffe},  i.e.,
$\epsilon^{\mu}\epsilon^{\nu}/\epsilon^2 \to \frac{1}{4}g^{\mu\nu}$, and the
fact that its second term contains only odd powers of $\epsilon$, we arrive at
\begin{equation}
\langle j^{\mu}_5 \rangle_{\rm singular} =
- \frac{\Delta}{2\pi^2\epsilon^2} \delta^{\mu}_{3} \sim
\frac{\Lambda^2 \Delta }{2\pi^2}\delta^{\mu}_{3} \,. \label{chiral-current}
\end{equation}
It is clear that $- 1/\epsilon^2$ plays the role of a Euclidean ultraviolet cutoff
$\Lambda^2$. This feature of expression (\ref{chiral-current}) agrees with the
results obtained in Ref.~\cite{Gorbar:2009bm}. As was shown there, while the
contribution of each Landau level into the axial current density $\langle
j^{\mu}_5 \rangle$ is finite at a fixed $\Delta$, their total contribution is
quadratically divergent. However, the important point is that since the
solution of gap equation (\ref{gap-Delta}) for the dynamical shift $\Delta$
yields $\Delta \sim g\mu\, eB/\Lambda^2 $, the axial current density is
actually finite. The explicit expressions for $\Delta$ and $\langle j^{3}_5
\rangle$ are \cite{Gorbar:2009bm}:
\begin{eqnarray}
\Delta &\simeq & -g \mu\frac{eB}{\Lambda^2\left(1+2 a g\right)},
\label{Delta-vs-mu}
\\
\langle j^{3}_5 \rangle
&\simeq & \frac{eB}{2\pi^{2}} \mu + a \frac{\Lambda^2}{\pi^{2}}  \Delta
\simeq \frac{eB}{2\pi^{2}}\frac{\mu}{\left(1+ 2 a g\right)},
\label{a}
\end{eqnarray}
where $a$ is a dimensionless constant of order one \cite{footnote1}, which is determined by the
regularization scheme used, and $g$ is the coupling constant defined in Sec.~\ref{2}. Note that 
both the topological and dynamical contributions are included in $\langle j^{3}_5 \rangle$. 
(Terms of higher order in powers of $|eB|/\Lambda^2$ are neglected in both expressions.)

In Ref.~\cite{Gorbar:2009bm}, a gauge noninvariant regularization (with a cutoff in
a sum over Landau levels) was used. One can show that the main features of the structure of the
axial current in model (\ref{NJLmodel}) remain the same also in the gauge invariant
proper-time regularization \cite{GMS}. In this regularization,
\begin{eqnarray}
\Delta &\simeq &  -g \mu \frac{eB}{\Lambda^2 (1+g/2)}, \\
\langle j_{3}^5 \rangle  &\simeq & \frac{eB}{2\pi^{2}} \mu +
\frac{\sqrt{\pi}}{2(2\pi l)^2}
\frac{e^{-s \Delta^2 }}{\sqrt{s}}
\mbox{erfi}(\sqrt{s} \Delta )\coth(eBs) \Bigg|_{s=1/\Lambda^2}
\simeq \frac{eB}{2\pi^{2}}\frac{\mu}{\left(1+g/2\right)},
\end{eqnarray}
where $\mbox{erfi}(x)\equiv -i\, \mbox{erf}(i x)$ is the imaginary error function. 
Note that in this case the parameter $a$ equals $1/4$ [compare with Eq.~(\ref{a})].

We conclude that interactions leading to the shift parameter $\Delta$ essentially change
the form of the induced axial current in a magnetic field. It is important to mention that
unlike the topological contribution in $\langle j^{\mu}_5 \rangle$ \cite{Metlitski:2005pr},
the dynamical one is generated by all Landau levels.

\subsection{Axial anomaly}
\label{3b}

In this subsection we will show that the shift parameter $\Delta$ does not affect
the axial anomaly.

In the gauge invariant point-splitting regularization, the divergence of the axial current
in the massless theory equals \cite{Peskin,Ioffe}
\begin{equation}
\partial_{\mu}j^{\mu}_5(u)=ie\epsilon^{\alpha}\bar{\psi}(u +\epsilon)
\gamma^{\mu}\gamma^5\psi(u)\left. F_{\alpha\mu}\right|_{ \epsilon \to 0}\,. 
\label{divergence}
\end{equation}
Then, calculating the vacuum expectation value of the divergence of the axial current, we find
\begin{equation}
\langle \partial_{\mu}j^{\mu}_5(u) \rangle
=-ie\epsilon^{\alpha}F_{\alpha\mu}\tr\left[\gamma^{\mu}\gamma^5
G(u,u+\epsilon) \right]_{\epsilon \to 0} =
ie\epsilon^{\alpha}F_{\alpha\mu} \langle j^{\mu}_5(u) \rangle\,,
\label{divergence1}
\end{equation}
where $G(u,u^{\prime})$ is the fermion propagator in Eq.~(\ref{propagator-transformed}).
Let us check that the presence of $\Delta$ in $G(u,u^{\prime})$, which modifies the axial
current, does not affect the standard expression for the axial anomaly.

We start by considering the first term in the axial current density in Eq.~(\ref{axialcurrent}):
\begin{equation}
\frac{i\epsilon^{\mu}s_{\perp}}{\pi^2\epsilon^4} \left(e^{- i
s_{\perp} \Delta \epsilon^3}-e^{i s_{\perp} \Delta \epsilon^3}\right) 
\simeq \frac{2\Delta\epsilon^{\mu}\epsilon^3}{\pi^2\epsilon^4}
\left(1-\frac{\Delta^2\epsilon^2_3}{6}+...\right)\,.
\label{chiral-current-series}
\end{equation}
Its contribution to the right-hand side of Eq.~(\ref{divergence1}) is
\begin{equation}
\frac{2i\Delta\,\epsilon^{\alpha}\epsilon^{\mu}\epsilon^3}{\pi^2\epsilon^4}
\left(1-\frac{\Delta^2\epsilon^2_3}{6}+...\right)eF_{\alpha \mu} \,.
\label{new-term}
\end{equation}
Since this expression contains only odd powers of $\epsilon$, it gives zero contribution after
symmetric averaging over space-time directions of $\epsilon$.

Thus, only the second term in  Eq.~(\ref{axialcurrent}) is relevant for the divergence of
axial current in Eq.~(\ref{divergence1}):
\begin{equation}
\langle \partial_{\mu}j^{\mu}_5(u) \rangle =
-\frac{e^2\epsilon^{\beta\mu\lambda\sigma}F_{\alpha\mu}F_{\lambda\sigma}
\epsilon^{\alpha}\epsilon_{\beta}}{8\pi^2\epsilon^2} \left(e^{- i s_{\perp} \Delta \epsilon^3}
+e^{i s_{\perp} \Delta \epsilon^3}\right) \to
-\frac{e^2}{16 \pi^2}
\epsilon^{\beta\mu\lambda\sigma}F_{\beta\mu}F_{\lambda\sigma}
\label{divergence-final}
\end{equation}
for $\epsilon\to 0$ and symmetric averaging over space-time directions of $\epsilon$. Therefore,
the presence of the shift parameter $\Delta$ does not affect the axial anomaly indeed.

\section{Discussion}
\label{discussion}

The emphasis in this Letter was on studying the structure of the induced axial 
current and the chiral anomaly in the normal phase in magnetized relativistic matter. 
Our conclusion is that there are two components in this current, the topological component,
induced only in the LLL, and the dynamical one provided by the chiral shift $\Delta$ 
(and generated in all Landau levels). While the former is intimately connected with
the axial anomaly, the latter does not affect the form of the anomaly at all. Thus
one can say that while the topological component of the current is anomalous, 
the dynamical one is normal.

The present analysis was realized in the NJL model. It would be important to extend 
it to renormalizable field theories, especially, QED and QCD. In connection with
that, we would like to note the following. The expression for the chiral shift parameter,
$\Delta \sim g\mu \, eB/\Lambda^2$, obtained in the NJL model implies that both fermion
density and magnetic field are necessary for the generation of $\Delta$. This feature
should also be valid in renormalizable theories. As for the cutoff $\Lambda$, it enters
the results only because of the nonrenormalizability of the NJL model. Similar studies of chiral
symmetry breaking in the vacuum ($\mu_0 = 0$) QED and QCD in a magnetic field show
that the cutoff scale $\Lambda$ is replaced by $\sqrt{|eB|}$ there \cite{QED}. Therefore,
one might expect that in QED and QCD with both $\mu$ and $B$ being nonzero, $\Lambda$
will be replaced by a physical parameter, such as $\sqrt{|eB|}$. This in turn suggests that
a constant chiral shift parameter $\Delta$ will become a running quantity that depends on the
longitudinal momentum $k^3$ and the Landau level index $n$.

It has been recently suggested in Refs.~\cite{Basar:2010zd,Kim}, that a chiral magnetic spiral
solution is realized in the chirally broken phase in the presence of a strong magnetic field. Like
the present solution with the chiral shift parameter $\Delta$, the chiral magnetic spiral one
is anisotropic, but beside that it is also inhomogeneous. It is essential, however, that the solution
with the chiral shift is realized in the {\it normal} phase of matter, in which the fermion density and
the axial current density are non-vanishing. It would be interesting to clarify whether there is a
connection between these two solutions describing the dynamics in the two different phases
of magnetized relativistic matter.

In this Letter, we concentrated on the basic and delicate questions regarding the chiral shift 
parameter $\Delta$, the induced axial current, and the axial anomaly, but did not address 
many specific details regarding the dynamics, e.g., those related to the chiral asymmetry 
of the Fermi surface \cite{Gorbar:2009bm} and a dependence of $\Delta$ on the temperature 
and the current fermion mass. These issues, which are of great interest because of their 
potential applications in neutron stars and in heavy ion collisions, will be considered 
elsewhere \cite{GMS}.

\begin{acknowledgments}
The authors would like to thank V.~P.~Gusynin for fruitful discussions. The work of E.V.G. was
supported partially by the SCOPES under Grant No. IZ73Z0-128026 of the Swiss NSF, under
Grant No. SIMTECH 246937 of the European FP7 program, the joint Grant RFFR-DFFD
No. F28.2/083 of the Russian Foundation for Fundamental Research and of the Ukrainian
State Foundation for Fundamental Research (DFFD). The work of V.A.M. was supported
by the Natural Sciences and Engineering Research Council of Canada. The work of I.A.S.
is supported in part by a start-up fund from the Arizona State University and by the U.S.
National Science Foundation under Grant No. PHY-0969844.
\end{acknowledgments}

\appendix
\section{Derivation of gap equation}
\label{AppPropagator}

In order to derive the gap equation, it is convenient to utilize the formalism of the effective
action for composite operators \cite{CJT}. In the mean field approximation that we use, the
corresponding effective action $\Gamma$ has the following form:
\begin{eqnarray}
\Gamma(G)&=& -i\mbox{Tr}\left[\mbox{Ln} G^{-1} +S^{-1}G-1\right]
+\frac{ G_{\rm int}}{2}\int dt \int d^{3}x\Big\{
\left(\tr\left[G(x,x)\right]\right)^{2}
-\left(\tr\left[\gamma^{5}G(x,x)\right]\right)^{2}
\nonumber\\
&&-\tr\left[G(x,x)G(x,x)\right]
+\tr\left[\gamma^{5}G(x,x)\gamma^{5}G(x,x)\right]\Big\},
\label{potential}
\end{eqnarray}
where the trace, the logarithm, and the product $S^{-1}G$ are taken in the functional sense.
Here $S$ and $G$ are the tree level fermion propagator and the full one, respectively. The free
energy density $\Omega$ is expressed through $\Gamma$ as $\Omega = -\Gamma/{\cal T}V$,
where ${\cal T}V$ is a space-time volume.

The stationarity condition $\delta\Gamma(G)/\delta{G}=0$ leads to the gap equation
\begin{eqnarray}
G^{-1}(u,u^\prime) &=& S^{-1}(u,u^\prime)
- i G_{\rm int} \Big\{ G(u,u) -  \gamma^5 G(u,u) \gamma^5
-  \tr[G(u,u)] +  \gamma^5\, \tr[\gamma^5G(u,u)]\Big\}
\delta^{4}(u- u^\prime).
\label{gap}
\end{eqnarray}
Here while the first two terms in the curly brackets describe the exchange (Fock) interactions, 
the last two terms describe the direct (Hartree) interactions.

The structure of $G^{-1}(u,u^\prime)$ is shown in Eq.~(\ref{ginverse}), and the structure of
the inverse tree level fermion propagator $S^{-1}(u,u^\prime)$ is determined from the Lagrangian
density in Eq.~(\ref{NJLmodel}):
\begin{eqnarray}
iS^{-1}(u,u^\prime) &=&\Big[(i\partial_t+\mu_0)\gamma^0
-(\bm{\pi}\cdot\bm{\gamma}) -\pi^3\gamma^3-m_{0}\Big]\delta^{4}(u- u^\prime).
\label{sinverse}
\end{eqnarray}
Comparing this expression with that in Eq.~(\ref{ginverse}), we see that the inverse full fermion
propagator $G^{-1}$ contains two new types of dynamical parameters that are absent at tree level
in $S^{-1}$: $\tilde{\mu}$ and $\Delta$. It is clear that $\tilde{\mu}$ plays the role of an anomalous
magnetic moment.

It should be emphasized that the Dirac mass and the chemical potential terms in the full propagator
are determined by $m$ and $\mu$ that may differ from their tree level counterparts, $m_0$ and
$\mu_0$. While $m_0$ is a bare fermion mass, $m$ is a constituent one, which in general depends
on the density and temperature of the matter, as well as on the strength of interactions. Concerning
the chemical potentials, it is $\mu_0$ that is the chemical potential in the thermodynamic sense.
The value of $\mu$, on the other hand, is an ``effective'' chemical potential that determines the
quasiparticle dispersion relations for fermion quasiparticles in interacting theory. As was already
pointed out in Sec. \ref{2}, Eq.~(\ref{gap-mu}) implies that at any nonzero fermion density
$\langle j^0 \rangle$, $\mu_0$ and $\mu$ are different if $G_{\rm int} \neq 0$.

In order to determine the values of the parameters $m$, $\mu$, $\Delta$ and $\tilde{\mu}$
in the model at hand, we use gap equation (\ref{gap}). As one can see, the right-hand side of
this equation depends only on the full fermion propagator $G(u,u^\prime)$ at $u^\prime=u$.
This fact greatly simplifies the analysis. Of course, it is related to the fact that we use the local
four-fermion interaction. This feature will be lost in more realistic models with long-range interactions.
The main disadvantage of the local four-fermion interaction is the nonrenormalizability of
the model. Therefore, model (\ref{NJLmodel}) should be viewed only as a low-energy effective
model reliable at the energy scales below a certain cutoff energy $\Lambda$.

The propagator $G(u,u)$ has the following Dirac structure:
\begin{equation}
G(u,u)= -\frac{1}{4}\left[\gamma^0{\cal A}+{\cal B}+ i\gamma^1\gamma^2{\cal C}+ \gamma^3\gamma^5{\cal D}\right].
\label{G(u,u)}
\end{equation}
The four coefficient functions can be defined through the following traces:
\begin{eqnarray}
 {\cal A} &=&-\tr\left[\gamma^0G(u,u)\right]
 \equiv \langle \bar\psi \gamma^0 \psi\rangle = \langle j^0 \rangle,
\label{A}\\
{\cal B}&=& -\tr\left[G(u,u)\right]
\equiv \langle \bar\psi \psi\rangle,
\label{B}\\
{\cal C}&=& -\tr\left[ i\gamma^1\gamma^2G(u,u)\right]
\equiv  \langle\bar\psi i\gamma^1\gamma^2\psi\rangle,
\label{C}\\
{\cal D} &=&-\tr\left[\gamma^3\gamma^5G(u,u)\right]
\equiv \langle \bar\psi \gamma^3\gamma^5\psi\rangle=  \langle j_5^3 \rangle.
\label{D}
\end{eqnarray}
Then the gap equation (\ref{gap}) can be rewritten in the following form:
\begin{equation}
(\mu-\mu_0)\gamma^0 + i\tilde{\mu}\gamma^1\gamma^2 + \Delta\gamma^3\gamma^5 -m+m_0 =
-\frac{1}{2}G_{\rm int}
\left[\gamma^0  {\cal A}+\gamma^3\gamma^5 {\cal D}\right] +G_{\rm int}  {\cal B}.
\label{gap-more}
\end{equation}
Note that function ${\cal C}$, related to the anomalous magnetic moment, see Eq.~(\ref{C}),
does not enter the right-hand side of the gap equation. Therefore, in the mean field approximation
used here, no nontrivial solution for $\tilde{\mu}$ is allowed \cite{footnote2}. The matrix equation
(\ref{gap-more}) with $\tilde{\mu} = 0$ is equivalent to the set of three algebraic
Eqs.~(\ref{gap-mu}) -- (\ref{gap-m}) in Sec.~\ref{2}.


\begin{thebibliography}{99}

\bibitem{astroreview}
P.~M.~Woods and C.~Thompson,
  ``Soft Gamma Repeaters and Anomalous X-ray Pulsars: Magnetar Candidates,''
  in {\it Compact Stellar X-ray Sources},
  edited by W. H. G. Lewin and M. van der Klis,
  (Cambridge University Press, Cambridge, 2006) pp. 547-586 [astro-ph/0406133].
  %%CITATION = ASTRO-PH/0406133;%%

\bibitem{astroreview1} S.~Mereghetti,
  %``The strongest cosmic magnets: Soft Gamma-ray Repeaters and Anomalous X-ray
  %Pulsars,''
  Astron.\ Astrophys.\ Rev.\  {\bf 15}, 225 (2008).
 %[arXiv:0804.0250 [astro-ph]].
  %%CITATION = AASRE,15,225;%%

\bibitem{Skokov:2009qp}
  V. V. Skokov, A. Yu. Illarionov and V. D. Toneev,
  %``Estimate of the magnetic field strength in heavy-ion collisions,''
  Int.\ J.\ Mod.\ Phys.\  A {\bf 24}, 5925 (2009).
  %arXiv:0907.1396 [nucl-th].
  %%CITATION = ARXIV:0907.1396;%%

\bibitem{Kharzeev:2007tn}
  D.~Kharzeev and A.~Zhitnitsky,
  %``Charge separation induced by P-odd bubbles in QCD matter,''
  Nucl.\ Phys.\  A {\bf 797}, 67 (2007);
  %%CITATION = NUPHA,A797,67;%%
%\bibitem{Kharzeev:2007jp}
  D.~E.~Kharzeev, L.~D.~McLerran, and H.~J.~Warringa,
  %``The effects of topological charge change in heavy ion collisions: 'Event by
  %event P and CP violation',''
  Nucl.\ Phys.\  A {\bf 803}, 227 (2008);
  %%CITATION = NUPHA,A803,227;%%
%\bibitem{Fukushima:2008xe}
  K.~Fukushima, D.~E.~Kharzeev, and H.~J.~Warringa,
  %``The Chiral Magnetic Effect,''
  Phys.\ Rev.\  D {\bf 78}, 074033 (2008);
  %%CITATION = PHRVA,D78,074033;%%
  Seung-il~Nam,
  %``Vector current correlation and charge separation via chiral magnetic
  %effect,''
  Phys.\ Rev.\  D {\bf 82}, 045017 (2010).
  %[arXiv:1004.3444 [hep-ph]].
  %%CITATION = PHRVA,D82,045017;%%

\bibitem{FI1}
  D.~Ebert, K.~G.~Klimenko, M.~A.~Vdovichenko, and A.~S.~Vshivtsev,
  %``Magnetic oscillations in dense cold quark matter with four-fermion
  %interactions,''
  Phys.\ Rev.\  D {\bf 61}, 025005 (1999);
  %% [arXiv:hep-ph/9905253].
  %%CITATION = PHRVA,D61,025005;%%
  E.~J.~Ferrer, V.~de la Incera, and C.~Manuel,
  %``Magnetic color flavor locking phase in high density QCD,''
  Phys.\ Rev.\ Lett.\  {\bf 95}, 152002 (2005);
  %%CITATION = PRLTA,95,152002;%%
  E.~J.~Ferrer and V.~de la Incera,
  %``Magnetic phases in three-flavor color superconductivity,''
  Phys.\ Rev.\  D {\bf 76}, 045011 (2007);
  %%CITATION = PHRVA,D76,045011;%%
 %``Color superconducting matter in a magnetic field,''
  Phys.\ Rev.\ Lett.\  {\bf 100}, 032007 (2008);
  %%CITATION = PRLTA,100,032007;%%
  J.~L.~Noronha and I.~A.~Shovkovy,
  %``Color-flavor locked superconductor in a magnetic field,''
  Phys.\ Rev.\  D {\bf 76}, 105030 (2007);
  %%CITATION = PHRVA,D76,105030;%%
  D.~T.~Son and M.~A.~Stephanov,
  %``Axial anomaly and magnetism of nuclear and quark matter,''
  Phys.\ Rev.\  D {\bf 77}, 014021 (2008).
  %%CITATION = PHRVA,D77,014021;%%

\bibitem{Metlitski:2005pr}
  M.~A.~Metlitski and A.~R.~Zhitnitsky,
  %``Anomalous axion interactions and topological currents in dense matter,''
  Phys.\ Rev.\  D {\bf 72}, 045011 (2005).
  %%CITATION = PHRVA,D72,045011;%%

\bibitem{Gorbar:2009bm}
  E.~V.~Gorbar, V.~A.~Miransky, and I.~A.~Shovkovy,
  %``Chiral asymmetry of the Fermi surface in dense relativistic matter in a
  %magnetic field,''
  Phys. Rev. C {\bf 80}, 032801(R) (2009).
  %%CITATION = PHRVA,C80,032801;%%

\bibitem{Rebhan}
  A.~Rebhan, A.~Schmitt and S.~A.~Stricker,
  %``Anomalies and the chiral magnetic effect in the Sakai-Sugimoto model,''
  JHEP {\bf 1001}, 026 (2010).
  %%CITATION = JHEPA,1001,026;%%

\bibitem{Fukushima}
  K.~Fukushima and M.~Ruggieri,
  %``Dielectric correction to the Chiral Magnetic Effect,''
  Phys. Rev. D {\bf 82}, 054001 (2010).
  %%CITATION = ARXIV:1004.2769;%%

\bibitem{Basar:2010zd}
  G.~Basar, G.~V.~Dunne and D.~E.~Kharzeev,
  %``Chiral Magnetic Spirals,''
  Phys.\ Rev.\ Lett.\  \textbf{104}, 232301 (2010).
  %%CITATION = PRLTA,104,232301;%%

\bibitem{Kim}
  K.~Y.~Kim, B.~Sahoo and H.~U.~Yee,
  %``Holographic chiral magnetic spiral,''
  JHEP {\bf 1010}, 005 (2010).
  %arXiv:1007.1985 [hep-th].
  %%CITATION = ARXIV:1007.1985;%%

\bibitem{Frolov}
  I.~E.~Frolov, K.~G.~Klimenko and V.~C.~Zhukovsky,
  %``Chiral density waves in quark matter within the Nambu--Jona-Lasinio model
  %in an external magnetic field,''
  Phys. Rev. D {\bf 82}, 076002 (2010).
  %%CITATION = ARXIV:1007.2984;%%

\bibitem{Rubakov}
  V.~A.~Rubakov,
  %``On chiral magnetic effect and holography,''
  arXiv:1005.1888 [hep-ph].
  %%CITATION = ARXIV:1005.1888;%%

\bibitem{ABJ}
  S.~L.~Adler,
  %``Axial vector vertex in spinor electrodynamics,''
  Phys.\ Rev.\  {\bf 177}, 2426 (1969);
  %%CITATION = PHRVA,177,2426;%%
    J.~S.~Bell and R.~Jackiw,
  %``A PCAC puzzle: pi0 $\to$ gamma gamma in the sigma model,''
  Nuovo Cimento  A {\bf 60}, 47 (1969).
  %%CITATION = NUCIA,A60,47;

\bibitem{Ambjorn}
The fact that in a magnetic field the axial anomaly is generated only in the LLL was shown in
J.~Ambjorn, J.~Greensite and C.~Peterson,
  %``The Axial Anomaly And The Lattice Dirac Sea,''
  Nucl.\ Phys.\  B {\bf 221}, 381 (1983);
  %%CITATION = NUPHA,B221,381;%%
N.~Sadooghi and A.~Jafari Salim,
  %``Axial anomaly of QED in a strong magnetic field and noncommutative
  %anomaly,''
  Phys.\ Rev.\  D {\bf 74}, 085032 (2006).
  %[arXiv:hep-th/0608112].
  %%CITATION = PHRVA,D74,085032;%%

\bibitem{footnote1}
Here our definition of the electric charge $e$ is such that $e=-|e|<0$ for the electron.

\bibitem{MC} V. P.~Gusynin, V. A.~Miransky, and I. A.~Shovkovy,
Phys. Rev. Lett. {\bf 73}, 3499 (1994);
%%CITATION = HEP-PH 9405262;%%
Phys. Lett. {\bf B349}, 477 (1995);
%%CITATION = PHLTA,B349,477;%%
Nucl.\ Phys.\ B {\bf 462}, 249 (1996).
%%CITATION = HEP-PH 9509320;%%

\bibitem{CJT}
  J.~M.~Cornwall, R.~Jackiw and E.~Tomboulis,
  %``Effective Action For Composite Operators,''
  Phys.\ Rev.\  D {\bf 10}, 2428 (1974).
  %%CITATION = PHRVA,D10,2428;%%

\bibitem{Kapusta}
  J.~I.~Kapusta,
  %``Bose-Einstein Condensation, Spontaneous Symmetry Breaking, And Gauge
  %Theories,''
  Phys.\ Rev.\  D {\bf 24}, 426 (1981);
  %%CITATION = PHRVA,D24,426;%%
V.~P.~Gusynin, V.~A.~Miransky and I.~A.~Shovkovy,
  %``Spontaneous rotational symmetry breaking and roton like excitations in
  %gauged sigma model at finite density,''
  Phys.\ Lett.\  B {\bf 581}, 82 (2004).
  %%CITATION = PHLTA,B581,82;%%

\bibitem{Peskin}   M.~E.~Peskin and D.~V.~Schroeder,
  ``An Introduction To Quantum Field Theory,''
%\href{http://www.slac.stanford.edu/spires/find/hep/www?irn=3485960}{SPIRES entry}
(Westview Press, 1995) p. 659.

\bibitem{Ioffe}
  B.~L.~Ioffe,
  %``Axial anomaly: The modern status,''
  Int.\ J.\ Mod.\ Phys.\  A {\bf 21}, 6249 (2006).
  %[arXiv:hep-ph/0611026].
  %%CITATION = IMPAE,A21,6249;%%
Recall that unlike the ordinary point-splitting regularization, a Wilson
line is inserted between the splitting points in the gauge invariant one.

\bibitem{Son}
  D.~T.~Son and A.~R.~Zhitnitsky,
  %``Quantum anomalies in dense matter,''
  Phys.\ Rev.\  D {\bf 70}, 074018 (2004).
  %%CITATION = PHRVA,D70,074018;%%

\bibitem{Gavai}
  R.~V.~Gavai and S.~Sharma,
  %``Anomalies at finite density and chiral fermions,''
  Phys.\ Rev.\  D {\bf 81}, 034501 (2010).
  %[arXiv:0906.5188 [hep-lat]].
  %%CITATION = PHRVA,D81,034501;%%

\bibitem{Schwinger}
  J.~S.~Schwinger,
  %``On gauge invariance and vacuum polarization,''
  Phys.\ Rev.\  {\bf 82}, 664 (1951).
  %%CITATION = PHRVA,82,664;%%

\bibitem{GMS}
E.~V.~Gorbar, V.~A.~Miransky, and I.~A.~Shovkovy
(in preparation).

\bibitem{QED}
V.~P.~Gusynin, V.~A.~Miransky and I.~A.~Shovkovy,
  %``Dynamical chiral symmetry breaking by a magnetic field in QED,''
  Phys.\ Rev.\  D {\bf 52}, 4747 (1995);
  %[arXiv:hep-ph/9501304].
  %%CITATION = PHRVA,D52,4747;%%
%``Theory of the magnetic catalysis of chiral symmetry breaking in QED,''
  Nucl.\ Phys.\  B {\bf 563}, 361 (1999);
  %[arXiv:hep-ph/9908320].
  %%CITATION = NUPHA,B563,361;%%
V.~A.~Miransky and I.~A.~Shovkovy,
  %``Magnetic catalysis and anisotropic confinement in QCD,''
  Phys.\ Rev.\  D {\bf 66}, 045006 (2002).
  %[arXiv:hep-ph/0205348].
  %%CITATION = PHRVA,D66,045006;%%

\bibitem{footnote2} The constraint $\tilde{\mu} = 0$
simplifies the analysis. However, we should emphasize that $\tilde{\mu}$ may well be non-vanishing
in more refined approximations and in models with other types of interactions
\cite{GGM2007,magCatFI}. Yet, as one learns from a similar analysis in graphene,
a nonzero $\tilde{\mu}$ should not change the main qualitative features of the phase with an
induced $\Delta$ \cite{GGM2007}.

\bibitem{GGM2007}
{E.~V.~Gorbar, V.P.~Gusynin, and V.~A.~Miransky,
Low Temp.\ Phys.\ {\bf 34}, 790 (2008);
%%CITATION = LTPHE,34,790;%%
E.~V.~Gorbar, V.~P.~Gusynin, V.~A.~Miransky, and I.~A.~Shovkovy,
  %``Dynamics in the quantum Hall effect and the phase diagram of graphene,''
  Phys.\ Rev.\  B {\bf 78}, 085437 (2008).}
  %%CITATION = PHRVA,B78,085437;%%

\bibitem{magCatFI}
E.~J.~Ferrer and V.~de la Incera,
%``Dynamically Induced Zeeman Effect in Massless QED,''
Phys. Rev. Lett. {\bf 102}, 050402 (2009).
%%CITATION = ARXIV:0807.4744;%%

\end{thebibliography}
\end{document}